\documentclass[prd,aps,floats,preprint,preprintnumbers,nofootinbib]{revtex4}
\usepackage{amsmath,amssymb,mathrsfs}
\usepackage{graphicx}
\usepackage{feynmf}
\usepackage{graphics}
\usepackage{amsmath,amscd}
\usepackage{epsfig}
\usepackage{dsfont}
\usepackage{latexsym,oldgerm}
\usepackage{color}
\def\beq{\begin{equation}}
\def\eeq{\end{equation}}
\def\bea{\begin{eqnarray}}
\def\eea{\end{eqnarray}}

\newcommand{\Az}{\mathcal{A}_0}
\newcommand{\A}{\mathcal{A}_{\theta}}
\newcommand{\AtMV}{\mathcal{A}_{\theta}^{\mathcal{M},V}}

\newcommand{\Dt}{\Delta_{\theta}}
\newcommand{\T}{{\rm T}}

\newcommand{\Pa}{\mathbb{C}\mathscr{P}}
\newcommand{\M}{\mathcal{M}}
\newcommand{\dx}{{\rm d}}
\newcommand{\e}{{\rm e}}

\newcommand{\I}{\mathds{1}}

\begin{document}
\small
\preprint{SU-4252-899 \vspace{1cm}} \setlength{\unitlength}{0.1mm}
\title{Inequivalence of QFT's on Noncommutative Spacetimes: Moyal versus Wick-Voros
\vspace{0.1cm}}
\author{ A. P.
Balachandran$^{a,b}$}\thanks{C\'atedra de Excelencia SANTANDER-UCIIIM\\$\ddagger$bal@phy.syr.edu}\author{A. Ibort$^b$}\thanks{albertoi@math.uc3m.es}\author{G. Marmo$^c$}\thanks{marmo@na.infn.it} \author{M. Martone$^{a,c}$}\thanks{mcmarton@syr.edu}
\affiliation{$^{a}$Department of Physics, Syracuse University, Syracuse, NY
13244-1130, USA\\
$^{b}$Departamento de Matem\'aticas, Universidad Carlos III de Madrid, 28911 Legan\'es, 
Madrid, Spain\\
$^{c}$Dipartimento di Scienze Fisiche, University of Napoli and INFN, Via Cinthia I-80126 Napoli, Italy}
\begin{abstract}
\vspace{0.1cm}

In this paper, we further develop the analysis started in an earlier paper on the inequivalence of certain quantum field theories on noncommutative spacetimes constructed using twisted fields. The issue is of physical importance. Thus it is well known that the commutation relations among spacetime coordinates, which define a noncommutative spacetime, do not constrain the deformation induced on the algebra of functions uniquely. Such deformations are all mathematically equivalent in a very precise sense. Here we show how this freedom at the level of deformations of the algebra of functions can fail on the quantum field theory side. In particular, quantum field theory on the Wick-Voros and Moyal planes are shown to be inequivalent in a few different ways. Thus quantum field theory calculations on these planes will lead to different physics even though the classical theories are equivalent. This result is reminiscent of chiral anomaly in gauge theories and has obvious physical consequences.

The construction of quantum field theories on the Wick-Voros plane has new features not encountered for quantum field theories on the Moyal plane. In fact it seems impossible to construct a quantum field theory on the Wick-Voros plane which satisfies all the properties needed of field theories on noncommutative spaces. The Moyal twist seems to have unique features which make it a preferred choice for the construction of a quantum field theory on a noncommutative spacetime.

\end{abstract}
\maketitle
\section{INTRODUCTION}\label{sec:intro}
It is a general belief that the structure of spacetime may change drastically at Plank scale. In particular in \cite{Doplicher} it has been shown, using general considerations on the coexistence of Einstein's theory of relativity and basic quantum physics, namely Heisenberg's uncertainty principle, that close to the Planck scale, spacetime may become noncommutative. We can model such
spacetime noncommutativity by the commutation relations
\beq \label{UV1}
[\widehat{x}_{\mu}, \widehat{x}_{\nu}] = i \theta_{\mu \nu} 
\eeq
where $\theta_{\mu \nu} = - \theta_{\nu \mu}$ are constants and 
$\widehat{x}_{\mu}$ are the coordinate functions on $\mathbb{R}^n$:
\beq
\widehat{x}_{\mu}(x) = x_{\mu}. 
\eeq

Relation (\ref{UV1}) can be implemented by deforming the product of the standard commutative algebra of functions $\Az\equiv(\mathscr{F}(\mathcal{M}),m_0$) on the Minkowski space-time ${\cal M}\cong \mathbb{R}^{4}$ into a noncommutative one. (Here $\mathscr{F}(\M)$ denotes smooth, complex valued, functions on $\M$). It has a unit which is preserved by deformation. The former one is a commutative algebra under the standard point-wise multiplication $m_0$:
\beq
m_{0} (f \otimes g)(x) = f(x)g(x).
\eeq

There is a general procedure to deform such a product in a controlled way using the so-called {\it twist deformation} \cite{drinfeld}. It consists in taking into account a limited set of noncommutative products, indicated by $m_\theta$, which can be related in a precise manner to the initial commutative one, $m_0$. The deformed algebra provided by the new product $m_\theta$ is named $\A\equiv(\mathscr{F}(\mathcal{M}),m_\theta)$. Specifically we only consider product $m_\theta$ of the form:
\beq\label{UV2}
m_{\theta}(f\otimes g)\equiv m_0\circ\mathcal{F}_{\theta}(f\otimes g)\quad,
\eeq
where $\mathcal{F}_\theta$ contains all the information on the ``amount of noncommutativity''. $\mathcal{F}_\theta$ is called the {\it twist} and is formally an invertible map from $\A\otimes\A\to\A\otimes\A$ whose dependence on the noncommutativity parameter $\theta$ is such that in the limit $\theta\to0$, $\mathcal{F}_\theta\to\I\otimes\I$. 

In the present paper we will only consider two particular choices for the twist and consequently for the multiplication map $m_\theta$: Moyal and Wick-Voros planes. The twists, multiplication maps and the deformed algebras of functions in the Moyal and Wick-Voros case will be respectively indicated by $(\mathcal{F}^\M_\theta,m^\M_\theta,\A^\M)$ and $(\mathcal{F}^V_\theta,m^V_\theta,\A^V)$. Both lead to (\ref{UV1}).

In the following, for the sake of simplicity, we will work in two dimensions. The generalization to arbitrary dimensions will be discussed in section VI.

In two dimensions, we can  always write $\theta_{\mu\nu}$ as
\bea\label{UV10}
\theta_{\mu\nu}&=&\theta\epsilon_{\mu\nu}\\
\epsilon_{01}=&-&\epsilon_{10}=1
\eea
where $\theta$ is a constant. Then the two twists $\mathcal{F}_{\theta}^{\mathcal{M},V}$ assume the form
\bea\label{UV8}
&\mathcal{F}^{\mathcal{M}}_{\theta}=\exp\frac{i}{2}\theta[\partial_x\otimes \partial_y-\partial_y\otimes \partial_x]\ ,&\\\label{UV9}
&\mathcal{F}^{V}_{\theta}=\exp\frac{1}{2}\theta[\partial_x\otimes \partial_x+\partial_y\otimes \partial_y]\mathcal{F}^{\mathcal{M}}_{\theta}=\mathcal{F}^\M_\theta\exp\frac{1}{2}\theta[\partial_x\otimes \partial_x+\partial_y\otimes \partial_y] \ .&
\eea

As can be easily checked, both $\A^{\mathcal{M}}$ and $\A^V$ realize the commutation relations (\ref{UV1}).  The noncommutative algebras of functions on spacetime with different twisted products which realize (\ref{UV1}) are in fact many more. As stated already above, hereafter we will only work with $\A^{\mathcal{M},V}$.

From (\ref{UV8}) and (\ref{UV9}) the noncommutative product on functions in the two cases follows immediately:
\bea
&&\qquad(f \star_{\mathcal{M}} g)(x)=m_{\theta} (f\otimes g)(x) = m_0\circ\mathcal{F}^{\mathcal{M}}_{\theta}(f \otimes g)(x)\equiv f(x)\textrm{e}^{\frac{i}{2}\theta_{\alpha \beta}\overleftarrow{\partial^{\alpha}}\otimes \overrightarrow{\partial^{\beta}}} g(x)\quad,\\\label{Voro}
&&(f \star_V g)(x)=m_{\theta} (f\otimes g)(x) = m_0\circ\mathcal{F}^V_{\theta}(f \otimes g)(x)\equiv f(x)\textrm{e}^{\frac{i}{2}\left(\theta_{\alpha \beta}\overleftarrow{\partial^{\alpha}}\otimes \overrightarrow{\partial^{\beta}} -i\theta\delta_{\alpha\beta}\overleftarrow{\partial^{\alpha}}\otimes \overrightarrow{\partial^{\beta}}\right)}g(x)\quad.
\eea
If we let the $\star$-product to act on the coordinate functions, we get in both cases the noncommutative relations (\ref{UV1}).

We may as well note here that just as two groups can be isomorphic, so too the algebras $\A^{\M,V}$ are isomorphic. That means that there exists an invertible linear map T$:\A^\M\to\A^V$ such that
\beq
{\rm T}(f\star_\M g)(x)=\Big[({\rm T}f)\star_V({\rm T}g)\Big](x)\quad.
\eeq
This T is given by
\beq
{\rm T}=\exp\left(-\frac{\theta}{4}\nabla^2\right)\quad.
\eeq
Note that T preserves conjugation, ${\rm T}(f^*)=({\rm T}f)^*$, so that T is a $*$-isomorphism.

From (\ref{UV1}), at first sight it seems that the noncommutativity of spacetime coordinates also violates Poincar\'e invariance: the L.H.S. of (\ref{UV1}) transforms in a non-trivial way under the standard action of the Poincar\'e group whereas the R.H.S. does not. The issue can be solved by noting that the L.H.S. of (\ref{UV1}) is to be interpreted in terms of tensor products and $m_\theta$:
\beq
[\hat{x}_\mu,\hat{x}_\nu]=m_\theta(\hat{x}_\mu\otimes\hat{x}_\nu-\hat{x}_\nu\otimes\hat{x}_\mu)\quad.
\eeq
The way the group acts on the tensor product space is a further information which is not given by the way elements of the group act on $\hat{x}_\mu$. For this we need to define a homorphism from $\mathscr{P}\to\mathscr{P}\otimes\mathscr{P}$ which takes the name of {\it coproduct} and is indicated by $\Delta$. (More precisely it is a homomorphism from the group algebra $\Pa$ to $\Pa\otimes\Pa$.) In physics the standard choice is the trivial map:
\beq\label{cop}
\Delta_0:g\in\mathscr{P}\to\Delta_0(g)=g\otimes g\in\mathscr{P}\otimes\mathscr{P}\quad.
\eeq
Once a group is provided with such a further structure $\Delta$, (fulfilling a few other requirements), we get what is called a Hopf algebra. It can then act on tensor products. For example, for $\Delta_0$, $\mathscr{P}$ acts on $\hat{x}_\mu\otimes\hat{x}_\nu$ according to $\hat{x}_\mu\otimes\hat{x}_\nu\to\Delta_0(g)\hat{x}_\mu\otimes\hat{x}_\nu:=(g\hat{x}_\mu\otimes g\hat{x}_\nu)$.

In \cite{chaichian,wess,sasha}, it has been shown that there exists a choice for $\Delta$, different from (\ref{cop}), which allows an action of the Poincar\'e group algebra (indicated in what follows by $g\triangleright f$) preserving the relations (\ref{UV1}). The new Poincar\'e action we get is called the {\it twisted action}. The coproduct  $\Delta_\theta$, which defines it, is called the {\it twisted coproduct}. Finally $\Delta_\theta$ changes the standard Hopf algebra structure associated with the Poincar\'e group (the Poincar\'e-Hopf algebra $H\mathscr{P}$) given by $\Delta_0$ (\ref{cop}) to a twisted Poincar\'e-Hopf algebra $H_{\theta}\mathscr{P}$. We now explain these twisted structures.

Following the notation used above, $H^{\M,V}_\theta\mathscr{P}$ and $\Delta^{\M,V}_\theta$ refer to the Moyal and Wick-Voros cases.

The explicit form for the deformation of $\Delta^{\M,V}_\theta$ can be obtained by asking the action of the Poincar\'e group to be an {\it automorphism} of the new algebra of functions $\A^{\mathcal{M},V}$ on spacetime. That is, the action of the group has to be compatible with the new noncommutative multiplication rule (\ref{UV2}):
\beq\label{UV5}
g\triangleright m_\theta^{\mathcal{M},V}(f\otimes h)(x)=m_\theta^{\mathcal{M},V}(g\triangleright (f\otimes h))(x)  \, .
\eeq

It is easy to see that the standard coproduct choice (\ref{cop}) which works for the commutative product $m_0$ is not compatible \cite{chaichian,wess,sasha} with the action of $\mathscr{P}$ on the deformed algebra $\AtMV$. In the cases under consideration, where $\A^{\mathcal{M},V}$ are twist deformations of $\mathcal{A}_0$, there is a simple rule to get deformations $\Dt^{\mathcal{M},V}$ of $\Delta_0$ compatible with $m_{\theta}$. They are given by the formula:
\beq\label{UV3}
\Dt^{\mathcal{M},V}=(F^{\mathcal{M},V}_{\theta})^{-1}\Delta_0 F^{\mathcal{M},V}_{\theta}
\eeq
where $F^{\mathcal{M},V}_{\theta}$ are elements in $H_{\theta}^{\mathcal{M},V}\mathscr{P}\otimes H_{\theta}^{\mathcal{M},V}\mathscr{P}$ and $\mathcal{F}_\theta^{\mathcal{M},V}$ are the corresponding realizations of the twist elements $F_{\theta}^{\mathcal{M},V}$ on $\AtMV$.

The deformations of $H\mathscr{P}=H_0\mathscr{P}$ we consider here are again very specific ones. We only change $\Delta_0$ to $\Dt^{\mathcal{M},V}$ leaving the group multiplication the same. For a deeper discussion on deformations of algebras and Hopf algebras, we refer again to the literature \cite{Dito,chari,majid,aschieri}. For the present work, we need just the essential ingredients for constructing a quantum field theory on noncommutative spacetimes. These are the deformed multiplication rules on the algebra of functions on spacetime (\ref{UV2}) and the consequent deformations of the co-product of the symmetry algebra $\Pa$ given by (\ref{UV3}). They modify the way in which $\mathscr{P}$ acts on tensor products and hence on multiparticle states.

The results discussed in this paper differ from those of \cite{Lizzi} because of the differences in the approaches in the construction of quantum field theories on $\A^{\M,V}$. In \cite{Lizzi} the existence of quantum field theories as in \cite{Vitale} on $\A^{\M,V}$ is assumed while here we will discuss their explicit construction.

\section{Weak Equivalence}

We already addressed the question of equivalence of two quantum field theories on noncommutative spaces in \cite{Mario}. We want to recall briefly here what we called ``classical equivalence'' there. 

Mathematically, in the theories we are dealing with, there are two deformations involved. The first one is at the product (algebraic) level because of the $\star$-product which makes the algebra of functions on spacetime noncommutative. The second is the Hopf algebraic deformation of the symmetry group acting on the deformed algebra of functions. We have shown in \cite{mario2} how the two are strongly tied, but still mathematically different.

Let us denote by $\mathcal{A}_\theta,H_\theta$ and $\mathcal{A}'_\theta,H'_\theta$ two different pairs of deformations of space-time and of the Hopf algebras of the kinematical group acting on them. We will say that the two theories constructed from them are ``weakly equivalent'' if both pair of algebras are equivalent $\mathcal{A}_\theta\cong\mathcal{A}'_\theta$ and $H_\theta\cong H'_\theta$, where the notion of equivalence of deformations of algebras and Hopf algebras can be found respectively in \cite{Dito} and \cite{majid}. (In \cite{Mario}, this equivalence was called ``classical equivalence'', but the new name seems more appropriate).

In \cite{Mario}, we have shown that if the pair of deformations are equivalent both at the algebraic and Hopf algebraic level, then the following diagram is commutative:
\begin{equation}\label{dia}
\begin{CD}
\A\otimes\A @>\mathrm{T}\otimes\mathrm{T}>> \A'\otimes\A'\\
@VV\Delta_{\theta}(g)V @VV\Delta'_{\theta}(\mathrm{T}g\mathrm{T}^{-1})V\\
\A\otimes\A @>\mathrm{T}\otimes\mathrm{T}>> \A'\otimes\A'
\end{CD}
\end{equation} 
for all $g\in H_\theta$.

Here the map T is the one which maps $\A$ to $\A'$ \cite{Mario}.  In section VI we will prove that if $\A\cong\A'$, then the two Hopf algebra deformations which are compatible with the product in each deformed algebra are also equivalent {\it provided} T $\in H_\theta$. This result reduces the ``weak equivalence'' of two field theories on noncommutative spacetimes to the requirement that the two algebras of functions are equivalent under the action of $H_\theta$.

The meaning of diagram (\ref{dia}) is simple. It is just the requirement that the map T which implements the isomporphism $\A\to\A'$ also correctly implements the isomorphism $H_\theta\to H'_\theta$.

We call (\ref{dia}) ``weak equivalence'' because (\ref{dia}) is not enough to establish the equivalence of quantum field theories on $H_\theta$ and  $H'_\theta$. We call the obstruction blocking the implementation of this weak equivalence in quantum field theories a ``quantum field anomaly''. It is discussed in what follows. It does not appear in quantum mechanics as already shown in \cite{Molina}.

\section{Quantum Field Theory on a noncommutative spacetime}

To proceed to a comparison of the two quantum field theories, namely Wick-Voros and Moyal, we should first  construct the former one. As already anticipated, in this construction, new features arise with respect to the standard quantum field theory on the Moyal plane \cite{sasha,mangano}. In this section, we are going to explain the general construction of a quantum field theory on a noncommutative spacetime. Then we will show that quantum field theory on the Wick-Voros spacetime is not consistent.

The twisted quantum fields should carry a unitary representation of the Poincar\'e group which implements the twisted coproduct. These fields should also implement the twisted statistics.

Let us first consider the Moyal case. As our previous work \cite{mangano} shows,
\bea\label{new1}
&a^\M_p=c_p\exp(-\frac{i}{2}p_\mu\theta^{\mu\nu}P_\nu)&\\\label{new2}
&a^{\M\dag}_p=c^\dag_p\exp(\frac{i}{2}p_\mu\theta^{\mu\nu}P_\nu)&
\eea
where $c_p$, $c^\dag_p$ are the untwisted $\theta^{\mu\nu}=0$ annihilation and creation operators. (We can assume all such operators to refer to in, out or free operators as the occasion demands), $p_\mu$ is the four momentum of the particle whereas $P_\mu$ is the momentum operator (of the fully interacting theory). If $(a,\Lambda)\to U(a,\Lambda)$ is the $\theta=0$ unitary representation of the Poincar\'e group, then these operators acting on the vacuum create states which transform with the Moyal coproduct under conjugation by $U(a,\Lambda)$ \cite{sasha,mangano}.

We remark that the Fock space we use here is ``standard'' and can be created by applying $c_p^\dag$'s on the vacuum. The unitarity of $U(a,\Lambda)$ is with regard to the scalar product on this Fock space.

Transformations  of the form (\ref{new1}) and (\ref{new2}) from $c_p$, $c^\dag_p$ to $a^\M_p$, $a^{\M\dag}_p$ appeared in the context of integrable models in 1+1 dimensions \cite{Grosse,Zamo,Faddeev} where they are called ``dressing transformations''. A discretised version of these formulas has in fact appeared there. For this reason, here too, we will call them dressing transformations.

In these equations, the dressing transformation could have been changed to 
\beq\label{Ago2}
a^{\mathcal{M}}_p=\exp\left({-\frac{i}{2}p_{\mu}\theta^{\mu\nu}P_{\nu}}\right)c_p
\eeq
\beq\label{Ago3}
a^{\mathcal{M}\dagger}_p=\exp\left({\frac{i}{2}p_{\mu}\theta^{\mu\nu}P_{\nu}}\right)c_p^{\dagger}
\eeq
But in fact (\ref{new1}) equals (\ref{Ago2}) and (\ref{new2}) equals (\ref{Ago3}) because $\theta^{\mu\nu}=-\theta^{\nu\mu}$ \cite{mangano,vaidya}. This observation is important. It ensures that $a_p^{\mathcal{M}}$ is the adjoint of $(a_p^{\mathcal{M}})^{\dagger}$ for the standard scalar product on Fock space.

Also the unitary representation of the Poincar\'e group, acting on untwisted operators, correctly reproduces the twisted transformation law \cite{sasha}.

One can deduce from (\ref{new1},\ref{new2}) that the twisted Moyal quantum field is
\beq\label{new3}
\varphi^\M_\theta=\varphi_0\e^{\frac{1}{2}\overleftarrow{\partial}_\mu\theta^{\mu\nu}P_\nu}\quad.
\eeq
This formula is first deduced for in, out or free fields. For example regarding (\ref{new1}) and (\ref{new2}) to refer to in operators and writing the in field in terms of $a_p^\M$, $a^{\M\dag}_p$ in the standard way,
\bea
&\varphi^{\M,{\rm in}}_\theta=\int\frac{\dx^3p}{2|p_0|}\big[a^\M_pe_{-p}+a^{\M\dag}_pe_p\big]\quad,&\\
&\e_p(x)=\e^{ip\cdot x},\quad\dx\mu(p):=\frac{\dx^3p}{2\sqrt{\vec{p}^2+m^2}}\quad m={\rm mass\ of\ the\ particle}\quad,&
\eea
we get, using (\ref{new1}) and (\ref{new2}),
\beq
\varphi^{\M,{\rm in}}_\theta=\varphi_0^{{\rm in}}\e^{\frac{1}{2}\overleftarrow{\partial}_\mu\theta^{\mu\nu}P_\nu}\quad.
\eeq
But since the interacting Heisenberg field becomes the in field as $x_0\to-\infty$,
\beq
\varphi_0(x)\to\varphi^{{\rm in}}_0\quad{\rm as}\quad x_0\to-\infty\quad,
\eeq
and $P_\mu$ is time-independent, we (at least heuristically) infer (\ref{new3}) for the fully interacting Heisenberg field.

An important feature of (\ref{new3}) is its self-reproducing property:
\beq\label{self1}
\underbrace{\varphi^\M_\theta\star_\M\varphi^\M_\theta\star_\M...\star_\M\varphi^\M_\theta}_{N-factors}=\varphi_0^N\e^{\frac{1}{2}\overleftarrow{\partial}_\mu\theta^{\mu\nu}P_\nu}
\eeq
This property plays a significant role in general theory. It is the basis for the proof of the absence of UV-IR mixing in Moyal field theories (with no gauge fields) \cite{BaPiQur,BaPiQue}.

Now consider the Wick-Voros case. The twisted creation operators which correctly create states from the vacuum transforming by the twisted coproduct are \cite{Mario}
\beq\label{new4}
a^{V\dag}_p=c^\dag_p\e^{\frac{i}{2}(p_\mu\theta^{\mu\nu}P_\nu-i\theta p_\nu P_\nu)}
\eeq
where $p_\nu P_\nu$ uses the Euclidean scalar product. Its adjoint is
\beq\label{new5}
a^V_p=\e^{-\frac{i}{2}(p_\mu\theta^{\mu\nu}P_\nu+i\theta p_\nu P_\nu)}c_p\quad.
\eeq
We prove elsewhere \cite{BaIbMaMa} that (\ref{new4}) and (\ref{new5}) are also dictated by the covariance of quantum fields.

The Moyal twist of $\varphi_0^\M$ is compatible with the adjointness operation since from (\ref{Ago2},\ref{Ago3}) we have for the adjoint ($a^{\M\dag}_p)^\dag$ of $a^{\M\dag}_p$,
\beq
(a^{\M\dag}_p)^\dag=\exp\left(-\frac{i}{2}p_\mu\theta^{\mu\nu}P_\nu\right)c_p=a^\M_p\quad.
\eeq
Thus we can put the dressing transformation on the right or on the left, and such flexibility is needed to preserve the $\dag$-operation: the dressed operator $a^\M_p$ is equal to the adjoint of the dressed operator $a^{\M\dag}_p$. This is the significance of the remark following (\ref{new1}-\ref{Ago3}).

The above property fails for the Wick-Voros case. Thus
\beq
a^V_p=c_p\exp\left(-\frac{i}{2}(p_\mu\theta^{\mu\nu}P_\nu+i\theta p_\nu P_\nu)-\frac{\theta}{2}p_\nu p_\nu\right)\neq c_p\exp\left(-\frac{i}{2}(p_\mu\theta^{\mu\nu}P_\nu+i\theta p_\nu P_\nu)\right)\quad.
\eeq
A consequence is that we have to twist the creation-annihilation parts $\varphi^{(\pm)I}_0$ (I=in, out or free) fields separately:
\bea\label{new13}
&\varphi_\theta^{(+)V,{\rm I}}=\int\dx\mu(p)a^{V,{\rm I}\dag}\e_p=\varphi_0^{(+){\rm I}}e^{\frac{1}{2}(\overleftarrow{\partial}_\mu\theta^{\mu\nu}P_\nu-i\theta\overleftarrow{\partial}_\mu P_\mu)}\quad,&\\
&\varphi_\theta^{(-)V,{\rm I}}=\int\dx\mu(p)a^{V,{\rm I}}\e_{-p}=e^{\frac{1}{2}(\overrightarrow{\partial}_\mu\theta^{\mu\nu}P_\nu+i\theta\overrightarrow{\partial}_\mu P_\mu)}\varphi_0^{(-){\rm I}}\quad,&\\
&\e_p(x)=\e^{ip\cdot x},\quad\dx\mu(p):=\frac{\dx^3p}{2\sqrt{\vec{p}^2+m^2}}\quad m={\rm mass\ of\ the\ field}\ \varphi_0^{\rm I}\quad,&
\eea
where now we have added the superscript I to $\varphi^{\rm I}_0$, $a^{V,{\rm I}\dag}_p$, and $a^{V,{\rm I}}_p$.

Therefore the field
\beq
\varphi^{V,{\rm I}}_\theta=\varphi^{(+)V,{\rm I}}_\theta+\varphi^{(-)V,{\rm I}}_\theta
\eeq
cannot be obtained by an overall twist acting on $\varphi^{\rm I}_0$. As we have to twist the creation and annihilation parts separately, we have to separately twist its positive and negative frequency parts $\varphi^{(\pm){\rm I}}_0$. But we cannot decompose the Heisenberg field $\varphi_0$ for $\theta^{\mu\nu}=0$ into $\varphi^{(\pm)}$ such that $\varphi_0^{(\pm)}\to\varphi^{(\pm){\rm I}}_0$ as $x_0\to\mp\infty$. That means that we do not know how to write the twisted Heisenberg field or develop the LSZ formalism for the Wick-Voros case. (The LSZ formalism for the Moyal case was developed from (\ref{new3}) in \cite{TRG}.)

But that is not all. The states created by the Wick-Voros quantum fields $\varphi^{(\pm)V}_\theta$ are not normalised in the same way as in the Moyal case. For instance
\bea\label{long3}
&&\quad\langle0|a^{V,{\rm I}}_{k_1}a^{V,{\rm I}}_{k_2}a^{V,{\rm I}\dag}_{p_2}a^{V,{\rm I}\dag}_{p_1}|0\rangle=\\\nonumber
&&\quad=\e^{\theta k_1\cdot k_2}4\sqrt{(\vec{k}_1^2+m^2)(\vec{k}_2^2+m^2)}\Big[\delta^3(k_1-p_1)\delta^3(k_2-p_2)+\e^{\frac{i}{2}k_{1\mu}\theta^{\mu\nu} k_{2\nu}}\delta^3(k_1-p_2)\delta^3(k_2-p_1)\Big],\\
&&\qquad\qquad\qquad\quad\qquad {\rm I=in,\ out,}\quad|0\rangle_{\rm in}=|0\rangle_{\rm out};\quad m={\rm mass\ of\ the\ field}\ \varphi_0^I\quad.
\eea

For scattering theory, normalisation is important. If we normalise the states as in the Moyal case, since the normalisation constant in (\ref{long3}) is momentum dependent, the normalised states no longer transform with the Wick-Voros coproduct.

The normalisation (\ref{long3}) has been computed using the standard scalar product in the Fock space. We can try changing it \cite{Mario} so that the states become correctly normalised. But then the representation $(a,\Lambda)\to U(a,\Lambda)$ ceases to be unitary.

A further point relates to the self-reproduction property of these Wick-Voros fields. $\varphi^{(\pm){\rm I}V}_\theta$ are self-reproductive, but in different ways. Thus
\bea
\underbrace{\varphi^{(+)V,{\rm I}}_\theta\star_V\varphi^{(+)V,{\rm I}}_\theta\star_V...\star_V\varphi^{(+)V,{\rm I}}_\theta}_{M-factors}=\Big(\varphi^{(+){\rm I}}_0\Big)^M\e^{\frac{1}{2}(\overleftarrow{\partial}_\mu)\theta^{\mu\nu}P_\nu-i\theta\overleftarrow{\partial}_\mu P_\mu}\\
\underbrace{\varphi^{(-)V,{\rm I}}_\theta\star_V\varphi^{(-)V,{\rm I}}_\theta\star_V...\star_V\varphi^{(-)V,{\rm I}}_\theta}_{M'-factors}=\e^{\frac{1}{2}(\overrightarrow{\partial}_\mu)\theta^{\mu\nu}P_\nu+i\theta\overrightarrow{\partial}_\mu P_\mu}\Big(\varphi^{(-){\rm I}}_0\Big)^{M'}
\eea
So $\varphi^{V,{\rm I}}_\theta$ does not have sellf-reproducing property as in (\ref{self1}).

\section{On a Similarity transformation}

There is no similarity transformation transforming $a^{\M,{\rm I}}_p$, $a^{\M,{\rm I}\dag}_p$, $a^{V,{\rm I}}_p$, $a^{V,{\rm I}\dag}_p$. One way to quickly see this is to examine the operators without the Moyal part of the twist. So we consider $c^{{\rm I}}_p$, $c^{{\rm I}\dag}_p$ and
\bea\label{new7}
a^{V,{\rm I}'}_p=\e^{\frac{1}{2}\theta p_\nu P_\nu}c^{\rm I}_p\quad,\\\label{new8}
a^{V,{\rm I}'\dag}_p=c^{{\rm I}\dag}_p\e^{\frac{1}{2}\theta p_\nu P_\nu}\quad.
\eea

Now
\bea
&[c^{\rm I}_p,c^{{\rm I}\dag}_k]=2|p_0|\delta^3(p-k)\I\quad,&\\
&p_0=\sqrt{\vec{p}^2+m^2},\quad m={\rm mass\ of\ the\ field}\ \varphi_0^{\rm I}\quad.&
\eea

If there existed a $W$ such that
\beq
Wc^{\rm I}_pW^{-1}=a^{V,{\rm I}'}_p,\quad Wc^{{\rm I}\dag}_pW^{-1}=a^{V,{\rm I}'\dag}_p\quad,
\eeq
then we would have
\beq
[a^{V,{\rm I}'}_p,a^{V,{\rm I}'\dag}_k]=2|p_0|\delta^3(p-k)\I\quad.
\eeq
But a direct calculation of the L.H.S. using (\ref{new7},\ref{new8}) shows that is is not equal to the R.H.S.. 

But there exists an $S$ which transforms $a^{\M,{\rm I}\dag}_p$ to $a^{V,{\rm I}\dag}_p$:
\bea\label{new9}
&S=\e^{\frac{\theta}{4}(P_\mu P_\mu+ K)}, \quad K=-\int\dx\mu(k)k_\mu k_\mu c^{{\rm I}\dag}_kc^{\rm I}_k&\\
&Sa^{\M,{\rm I}\dag}_pS^{-1}=a^{V,{\rm I}\dag}_p\quad.&
\eea
where, as usual, I on $c^{{\rm I}\dag}_k$, $c^{\rm I}_k$ denotes in, out or free while in $P_\mu P_\mu$ and $k_\mu k_\mu$ we use the Euclidean scalar product.

But
\beq
Sa^{\M,{\rm I}}_pS^{-1}=\e^{-\frac{i}{2}(p_\mu\theta^{\mu\nu}P_\nu-i\theta p_\nu p_\nu)}c^{\rm I}_p=\tilde{a}^{V,{\rm I}}_p\neq a^{V,{\rm I}}_p\quad.
\eeq

Let us pursue the properties of this operator further.

The operator $S$ leaves the vacuum invariant and shows that certain correlators in the Moyal and Wick-Voros cases are equal. From the explicit expression (\ref{new9}) follows also that the map induced by the operator $S$ is isospectral, but not unitary in the standard Fock space scalar product. It is possible to define a new scalar product which makes $S$ unitary \cite{Mario}. But $U(a,
\Lambda)$ is not unitary in this scalar product.

Now consider the twisted fields
\bea
\varphi^{\M,{\rm I}}_\theta=\int\dx\mu(p)\big[a^{\M,{\rm I}}_p\e_{-p}+a^{\M,{\rm I}\dag}_p\e_{p}\big]\quad,\\
\tilde{\varphi}^{V,{\rm I}}_\theta=\int\dx\mu(p)\big[\tilde{a}^{V,{\rm I}}_p\e_{-p}+a^{V,{\rm I}\dag}_p\e_{p}\big]\quad,
\eea
where $\e_p(x)$ denotes as usual $\e^{ip\cdot x}$. Then of course,
\beq\label{new10}
S:\varphi^{\M,{\rm I}}_\theta\to S\triangleright\varphi^{\M,{\rm I}}_\theta:=\int\dx\mu(p)S\big[a^{\M,{\rm I}}_p\e_{-p}+a^{\M,{\rm I}\dag}_p\e_p\big]S^{-1}=\tilde{\varphi}^{V,{\rm I}}_\theta\quad.
\eeq
Also
\beq\label{new11}
S|0\rangle=S^{-1}|0\rangle=0
\eeq

From (\ref{new10}) and (\ref{new11}) we obtain trivially the equality of the $n$-points correlation functions:
\beq
\langle\varphi^{\M,{\rm I}}_\theta(x_1)\varphi^{\M,{\rm I}}_\theta(x_2)...\varphi^{\M,{\rm I}}_\theta(x_N)\rangle_0=\langle\tilde{\varphi}^{V,{\rm I}}_\theta(x_1)\tilde{\varphi}^{V,{\rm I}}_\theta(x_2)...\tilde{\varphi}^{V,{\rm I}}_\theta(x_N)\rangle_0\quad.
\eeq

Consider simple interaction densities such as
\beq
\mathscr{H}^\M_{\rm I}=\underbrace{\varphi^{\M,{\rm I}}_\theta\star_\M\varphi^{\M,{\rm I}}_\theta\star_\M...\star_\M\varphi^{\M,{\rm I}}_\theta}_{N-factors}\quad{\rm and}\quad\mathscr{H}^V_{\rm I}=\underbrace{\tilde{\varphi}^{V,{\rm I}}_\theta\star_V\tilde{\varphi}^{V,{\rm I}}_\theta\star_V...\star_V\tilde{\varphi}^{V,{\rm I}}_\theta}_{N-factors}\quad.
\eeq
in either fields.

Since $S$ only acts on the operator parts of the fields, the similarity transformation in (\ref{new10}) will not map $\mathscr{H}^\M_{\rm I}$ to $\mathscr{H}^V_{\rm I}$:
\beq
S\triangleright\mathscr{H}^\M_{\rm I}\neq\mathscr{H}_{\rm I}^V\quad.
\eeq
Hence
\bea
&&\langle\varphi^{\M,{\rm I}}_\theta(x_1)...\varphi^{\M,{\rm I}}_\theta(x_j)\mathscr{H}^\M_{\rm I}(x_{j+1})\varphi^{\M,{\rm I}}_\theta(x_{j+2})...\varphi^{\M,{\rm I}}_\theta(x_N)\rangle_0\\
&&\qquad\qquad\qquad\qquad\neq\langle\tilde{\varphi}^{V,{\rm I}}_\theta(x_1)...\tilde{\varphi}^{V,{\rm I}}_\theta(x_j)\mathscr{H}^V_{\rm I}(x_{j+1})...\tilde{\varphi}^{V,{\rm I}}_\theta(x_{j+2})...\tilde{\varphi}^{V,{\rm I}}_\theta(x_N)\rangle_0\quad.
\eea
So we can immediately conclude that also in this case the two theories are different.

There is no such $S$ for mapping $\varphi^{\M,{\rm I}}_\theta$ to $\varphi^{V,{\rm I}}_\theta$, so that the correlators are not equal even at the free level.

\section{a criterion for the strong equivalence of Twisted QFT's}

It seems reasonable to assert that two twisted quantum field theories obtained by twisting the same quantum field $\varphi_0$ are {\it strongly} equivalent if they give the same answer for the same scattering cross sections. This criterion is logically distinct from the criterion requiring the equality of Wightman functions, but is perhaps physically more compelling. The reason that the equality of Wightman functions and that of scattering cross sections need not mutually imply each other is the following. Below, in (\ref{Sca1}) and (\ref{Sca11}), we have given the scattering amplitudes in the Moyal and Wick-Voros cases. Even if they were equal due to equality of Wightman functions, it does not mean that the corresponding cross sections are equal, as the states in the two cases are not normalised in the same way.

Let us first recall the expression for a general scattering amplitude of spinless particles of mass $m_i$ in the Moyal case using the  LSZ formalism.

As argued heuristically in \cite{TRG}, the LSZ formalism for the Moyal field $\varphi^\M_\theta$ leads to the scattering amplitude
\beq\label{Sca1}
S_\theta^\M(k_1,...,k_N)= \langle-k_M,-k_{M-1},...,-k_1;{\rm out} \, | \, k_{N},k_{N-1},...,k_{N-M};{\rm in}\rangle_\M=\int\mathcal{I}\ G^\M_{N}(x_1,x_2,...,x_{N})
\eeq
where
\bea\label{Sca11}
&G_N^\M(x_1,...,x_N)=T\e^{\frac{i}{2}\sum_{I<J}\partial_I\wedge\partial_J}W^0_N(x_1,...,x_N)=T\ W^\M_N(x_1,...,x_N)&\quad,\\
&\mathcal{I}=\prod_{i=1}^N\dx x_i\e^{-iq_i\cdot x_i}i(\partial_i^2+m^2)&\quad.
\eea
The momenta $k_i$ are taken to be in-going so that $\sum k_i=0$. Also since
\beq
a_k^{\M\dag}|0\rangle=c^\dag_k|0\rangle\quad,
\eeq
the single particle states are normalised canonically:
\beq
\langle0|a^\M_{k'}a^{\M\dag}_k|0\rangle=2|k_0|\delta^3(k-k')
\eeq
while the normalisation of the multiparticle states
\beq
a^{\M\dag}_{k_1} \cdots a^{\M\dag}_{k_N}|0\rangle
\eeq
is consistent with what is required by twisted statistics.

For the Wick-Voros case, we can tentatively construct an in, out or free Wick-Voros field $\varphi^{V,{\rm I}''}_\theta$ following the construction (\ref{new13}) of $\varphi^{(+)V,{\rm I}}_\theta$:
\beq\label{new12}
\varphi^{V,{\rm I}''}_\theta=\varphi^{\rm I}_0(x)\e^{\frac{1}{2}\big(\overleftarrow{\partial}_\mu\theta^{\mu\nu}P_\nu-i\theta\overleftarrow{\partial}_\mu P_\mu\big)}\quad.
\eeq
The annihilation part of the $\varphi^{V,{\rm I}''}_\theta$ differs from $\varphi^{(-)V,{\rm I}}_\theta$ so that $\varphi^{V,{\rm I}''}_\theta$ does not have correct adjointness properties. But the formula (\ref{new12}) does generalise to Heisenberg fields. Using (\ref{new12}), we can obtain a formula like (\ref{Sca1}) for scattering amplitudes. It is
\bea
&S_\theta^{V''}(k_1,...,k_N)= \langle-k_M,-k_{M-1},...,-k_1;{\rm out}\, | \, k_{N},k_{N-1},...,k_{N-M};{\rm in}\rangle_V=\int\mathcal{I}G^{V''}_{N}(x_1,...,x_{N}),&\\
&\qquad G_N^{V''}(x_1,...,x_N)=T\e^{\frac{i}{2}\sum_{I<J}\partial_I\wedge\partial_J}\e^{\frac{\theta}{2}\sum_{I<J}\partial_I\cdot\partial_J}W^0_N(x_1,...,x_N)=T\ W^V_N(x_1,...,x_N)&
\eea
where $\partial_I\cdot\partial_J$ uses the Euclidean scalar product.

There is no reason to expect that $S_\theta^{V''}(k_1,...,k_N)=S_\theta^\M(k_1,...,k_N)$. In particular there is a problem with the normalisation of the states associated with $a^{V,{\rm I}\dag}$ as was pointed out already in (\ref{long3}).

We note however that the field (\ref{new12}) does have the self-reproducing property.

\section{Further Remarks on weak equivalence}

We can now briefly outline how to generalize our considerations on the Wick-Voros twist (\ref{UV9}) to $2N$-dimensions\footnote{In 2$N+1$-dimensions, we can always choose $\theta_{\mu\nu}$ so that $\theta_{\mu,2N+1}$=$\theta_{2N+1,\mu}$=0.}. We can always choose $\hat{x}_{\mu}$ so that $\theta_{\mu\nu}$, now an $2N\times 2N$ skewsymmetric matrix, becomes a direct sum of $N$ $2\times 2$ ones.  These $2\times2$ matrices are of the form (\ref{UV10}), but different 2$\times$2 matrices may have different factors $\theta$. For every such $2\times2$ block, we have a pair of $\hat{x}$'s which can be treated as in the 2-dimensional case above. (Of course there is no twist in any block with a vanishing $\theta$.)

We want also to show the technical result anticipated in section III which connects the cohomologies describing deformations of Hopf algebras and their module algebras. In particular, given two deformations of the spacetime algebra of functions $\A$ and $\A'$ and the compatible deformations of the action of Poincar\'e group algebra $\Pa_\theta$ and $\Pa'_\theta$ in the sense of (\ref{UV3}), the condition of equivalence of two algebraic deformations is:
\beq\label{equi}
\forall \ f_1,f_2\in\A,\quad\T\Big[(f_1\star f_2)\Big](x)=\Big[\T(f_1)\star'\T(f_2)\Big](x)\quad.
\eeq
where $\star$ and $\star'$ are the deformed products in, respectively, $\A$ and $\A'$ and $\T:\A\to\A'$ is the invertible map which implements the equivalence. 

The condition of equivalence of the two Hopf algebras characterized by the twists $F_\theta$ and $F'_\theta$, call them $H\mathscr{P}_\theta$ and $H'\mathscr{P}_\theta$ \cite{majid}, is
\beq\label{Mar6}
F'_\theta=\Delta_0({\rm T'})F_\theta{\rm T'}^{-1}\otimes {\rm T'}^{-1}\quad.
\eeq
Here T$'$ maps $H$ to $H'$ according to
\beq
H\ni h\to \mathrm{T'}h\mathrm{T'}^{-1}=h'\in H'
\eeq
where T$'$ is an element of the Hopf algebra.
The condition of ``weak equivalence'' also involves a further condition: the map T and T$'$ must be the same.

This condition of weak equivalence can be formulated for any two Hopf algebras $H$ and $H'$ acting on two algebras $\mathcal{A}$ and $\mathcal{A}'$ respectively if the following conditions are fulfilled:
\begin{itemize}
\item[1)] $\mathcal{A}$ and $\mathcal{A}'$ as vector (topological) spaces are the same and differ only in their multiplications maps $m$ and $m'$.

\item[2)] The Hopf algebras $H$ and $H'$ as algebras are the same and act on elements of $\mathcal{A}$ and $\mathcal{A}'$ in the same way. They differ only in their coproducts.

\item[3)] The products $m$ and $m'$ in $\mathcal{A}$ and $\mathcal{A}'$ are given by twists $\mathcal{F}\in H\otimes H$ and $\mathcal{F}'\in H'\otimes H'$ and a common multiplication map $m_0$ as follows:
\bea
m(f\otimes g)=m_0\mathcal{F}(f\otimes g)\\
m'(f\otimes g)=m_0\mathcal{F}'(f\otimes g)
\eea

\item[4)] The algebra $\mathcal{A}_0$ with the multiplication map $m_0$ is also a module for a Hopf algebra $H_0$. $H_0$ differs from $H$ and $H'$ only in its coproduct. It acts on elements of $\mathcal{A}_0$ just as $H$ and $H'$ act on elements of $\mathcal{A}$ and $\mathcal{A}'$

\end{itemize}

We consider only such algebras below. They cover the case of Moyal and Wick-Voros algebras and their corresponding Poincar\'e-Hopf algebras.

We are now going to show that if the two algebras $\mathcal{A}$ and $\mathcal{A}'$ are equivalent, that is (\ref{equi}) is satisfied, then the equivalence automatically lifts to the equivalence of the corresponding Hopf algebras {\it provided that} ${\rm T}\in H$. In that case ${\rm T'=T}$ We can hence say that what has been called ``weak equivalence'' is nothing but the equivalence of the two algebras $\mathcal{A}$ and $\mathcal{A}'$ under a map T$\in H$.

Let $\Delta_0$, $\Delta$ and $\Delta'$ be the coproducts for $H_0$, $H$ and $H'$. The proof is easily obtained by writing (\ref{equi}) using (\ref{UV2}):
\beq
{\rm T}\Big[m_0\circ\mathcal{F}(f_1\otimes f_2)\Big](x)=m_0\circ\mathcal{F}'\Big[{\rm T}\otimes{\rm T}(f_1\otimes f_2)\Big](x)\quad.
\eeq
Since the co-product compatible with the point-wise product is $\Delta_0$, we get:
\beq
\Big[m_0\circ\Delta_0({\rm T})\mathcal{F}_\theta(f_1\otimes f_2)\Big](x)=m_0\circ\mathcal{F}'_\theta\Big[{\rm T}\otimes{\rm T}(f_1\otimes f_2)\Big](x)
\eeq
which translates exactly into the equivalence condition (\ref{Mar6}) on the twists.

\section{Conclusions}

In the present paper we have addressed the question of how the freedom left by relations (\ref{UV1}) on possible algebra and Hopf algebra deformations implementing them, gets reflected on the quantum field theory side. The study was started in \cite{Mario}, but in the present work, a much more systematic presentation has been given. The approach to noncommutative quantum field theories we have used throughout the paper is based on the twisted fields \cite{sasha}. Our analysis shows that although the different algebra and Hopf algebra deformations may belong to the same equivalence class, the process of quantization introduces nontrivial subtleties which deserve to be studied in more depth in the future. Specifically we dealt with two twist deformations leading to the weakly equivalent Moyal and Wick-Voros algebras of functions. We then showed that quantum field theories on these planes are inequivalent. In particular the necessity of a non-unitary term in the dressing transformation of quantum fields on the Wick-Voros plane seems to make quantum field theories on this plane inconsistent. This result gives a preferred status to the Moyal twist to construct noncommutative quantum field theories. 

The issues we have addressed here must be studied further. In particular a clearer physical understanding of the freedom available in choosing a particular algebra and Hopf algebra deformation is needed. Thus in noncommutative geometry we take commutation relations among spacetime coordinates such as (\ref{UV1}) to be fundamental to define the noncommutativity of spacetime, the particular twist deformations like (\ref{UV8}) and (\ref{UV9}) being just ways to implement them. The inequivalence of quantum field theories derived from different algebra deformations raises questions about such a point of view.

\section{Acknowledgements}

It is a pleasure for Balachandran, Marmo and Martone to thank Alberto Ibort and the Universidad Carlos III de Madrid for their wonderful hospitality and support. It is a pleasure for Balachandran and Martone also to thank Alvaro Ferraz and the Centro Internacional de F\'isica da Mat\'eria Condensada of Brasilia where this work started. We also thank A. Pinzul and A. R. Queiroz for extensive and very useful discussions at Brasilia about aspects of this work at its initial stages.

The work of Balachandran and Martone was supported in part by DOE under the grant number DE-FG02-85ER40231 and by the Department of Science and Technology, India.


\bibliographystyle{apsrmp}

\end{document}